\UseRawInputEncoding
\documentclass[superscriptaddress, twocolumn, amsmath, amssymb, aps,prl, letterdraft, notitlepage,longbibliography]{revtex4-1}
\usepackage{graphicx,graphics,epsfig,subfigure,times,bm,bbm,amssymb,amsmath,amsfonts,amsthm,mathrsfs,MnSymbol}
\usepackage[matrix,frame,arrow]{xypic}
\usepackage[normalem]{ulem}
\usepackage{slashed}
\usepackage{color}
\usepackage[usenames,dvipsnames,svgnames,table]{xcolor}
\usepackage[english]{babel}
\usepackage{verbatim}
\usepackage{tabularx}
\definecolor{darkblue}{rgb}{0.0,0.0,0.3}
\usepackage[colorlinks=true,
            linkcolor=red,
            urlcolor= darkblue,
            citecolor=blue]{hyperref}

\newcommand{\bea}{\begin{eqnarray}}
\newcommand{\eea}{\end{eqnarray}}

\begin{document}
%\title{Strong coupling hinders optimal performance of linear cyclic quantum heat engines}
\title{Optimal linear cyclic quantum heat engines cannot benefit from strong coupling}

\author{Junjie Liu}
\email{jj\_liu@shu.edu.cn}
\address{Department of Physics, International Center of Quantum and Molecular Structures, Shanghai University, Shanghai, 200444, China}
\author{Kenneth A. Jung}
\email{kajung@stanford.edu}
\address{Department of Chemistry, Stanford University, Stanford, California, 94305, USA}

\begin{abstract}
Uncovering whether strong system-bath coupling can be an advantageous operation resource for energy conversion can facilitate the development of efficient quantum heat engines (QHEs). Yet, a consensus on this ongoing debate is still lacking owing to challenges arising from treating strong couplings. Here we conclude the debate for optimal linear cyclic QHEs operated under a small temperature difference by revealing the detrimental role of strong system-bath coupling in their optimal operations. We analytically demonstrate that both the efficiency at maximum power and maximum efficiency of strong-coupling linear cyclic QHEs are upper bounded by their weak-coupling counterparts and, particularly, experience a quadratic suppression relative to the Carnot limit under strong time-reversal symmetry breaking. 
\end{abstract}

\date{\today}

\maketitle

{\it Introduction.--} 
The miniaturization of controllable quantum systems opens doors for realizing nanoscale quantum heat engines (QHEs) that enable heat-work conversion in the quantum realm \cite{Abah.12.PRL,Rossnagel.14.PRL,Dechant.15.PRL,Robnagel.16.S,Sood.16.NP,Peterson.19.PRL,Klatzow.19.PRL,Assis.19.PRL,Pekola.19.ARCMP,Lindenfels.19.PRL}. At the nanoscale, the surface area of the working substances of QHEs could easily become comparable to their volume, which gives rise to scenarios where the strong system-bath coupling limit is attainable. Investigating such strong-coupling QHEs requires a quantum thermodynamic framework that extends beyond the classical version where system-bath coupling is assumed to be negligible \cite{Pachn2019}. Hence strong-coupling QHEs can serve as a vital platform for demonstrating intrinsic quantum signatures of energy conversion \cite{Uzdin.15.PRX,Klatzow.19.PRL,Dann.20.NJP}. Moreover, analyzing the performance of strong-coupling QHEs allows for validation of proposed definitions for thermodynamic quantities at strong couplings \cite{Rivas.20.PRL,Strasberg.20.PRE}, an ongoing topic of strong-coupling thermodynamics (see a recent review \cite{Talkner.20.RMP} and references therein).

Understanding the role of system-bath coupling in heat-work conversion can advance the field of strong-coupling QHEs. Substantial efforts have been put into the investigation of 
whether strong system-bath coupling can lead to operation advantages. To date, no general consensus has yet been reached, in part due to theoretical and numerical challenges imposed by strong system-bath couplings \cite{Talkner.20.RMP}.
In this ongoing debate there are studies that claim system-bath coupling could be a useful resource which potentially enhances the performance of QHEs \cite{Klimovsky.15.JPCL,Strasberg.16.NJP,Xu.18.PRE,Liu.21.PRL,Shirai.21.PRR,Carrega.22.PRXQ} and there are results suggesting detrimental effects of finite system-bath coupling \cite{Gallego.14.NJP,Kato.16.JCP,Newman.17.PRE,Restrepo.18.NJP,Newman.20.PRE,Perarnau-Llobet.18.PRL,Kaneyasu.22.A}. 
This lack of agreement stems from the fact that existing studies on strong-coupling QHEs are largely carried out on either specific models \cite{Klimovsky.15.JPCL,Kato.16.JCP,Strasberg.16.NJP,Newman.17.PRE,Restrepo.18.NJP,Newman.20.PRE,Carrega.22.PRXQ} or specific cycles \cite{Gallego.14.NJP,Perarnau-Llobet.18.PRL,Xu.18.PRE,Liu.21.PRL,Shirai.21.PRR,Kaneyasu.22.A} which limits the generality of their conclusions on the role of system-bath coupling in heat-work conversion. 

Here, we focus on {\it generic} periodically driven QHEs from weak to strong couplings operated in the linear response regime characterized by a small temperature difference (we will refer to these as linear cyclic QHEs hereafter). Using a complete form of the first law of thermodynamics which holds for generic cyclic QHEs \cite{Liu.21.PRL} and leveraging the principals of linear irreversible thermodynamics \cite{Callen.85.NULL}, we reveal a {\it universal} feature of linear cyclic QHEs that {\it optimal} weak-coupling machines perform more efficiently than their strong-coupling counterparts with the same degree of time-reversal symmetry breaking, conditional only on the non-negativity of both the entropy production rate and the efficiency of QHEs. We gain this general insight by first obtaining thermodynamic bounds on the efficiency at maximum power and maximum efficiency [cf. Eqs. (\ref{eq:eta_P_max}) and (\ref{eq:eta_max})] of linear cyclic QHEs valid from weak to strong couplings, then demonstrating that both the efficiency at maximum power and maximum efficiency of linear cyclic QHEs are upper bounded by their weak-coupling limits [cf. Eqs. (\ref{eq:eta_lim}) and (\ref{eq:eta_m_lim})].
Our thermodynamic bounds on optimal figures of merit reduce to known forms \cite{Benenti.11.PRL,Jiang.14.PRE} in the weak coupling limit, thereby indicating that the existing thermodynamic bounds \cite{Benenti.11.PRL,Jiang.14.PRE} when applying to cyclic QHEs are only applicable at weak couplings \cite{Brandner.15.PRX,Brandner.16.PRE}. We also find that both the efficiency at maximum power and maximum efficiency of strong-coupling linear cyclic QHEs are quadratically suppressed from the Carnot limit under strong time-reversal symmetry breaking [cf. Eqs. (\ref{eq:eta_lim}) and (\ref{eq:eta_m_lim})]. Our findings uncover a universal detrimental role of strong system-bath coupling in shaping the optimal performance of generic linear cyclic QHEs, and provide crucial insights into the search of efficient strong-coupling QHEs over weak-coupling counterparts as one steps out of the linear response regime. 

{\it Linear cyclic QHEs.--}
We consider generic cyclic QHEs as described by the total Hamiltonian ($\hbar=1$ and $k_B=1$ hereafter)
\begin{equation}\label{eq:HH}
H(t)~=~H_S(t)+H_I(t)+H_B.
\end{equation}
Here, $H_S(t)$ describes a periodically-driven working substance, $H_B=\sum_{v=h,c}H_B^v$ includes a hot (h) and a cold (c) heat bath at temperatures $T_v$. $H_I(t)=\sum_v H_I^v(t)$ denotes a time-dependent system-bath coupling allowing for the implementation of thermodynamic strokes. We take periodic protocols such that $H_{S,I}(\mathcal{T})=H_{S,I}(0)$ with $\mathcal{T}$ being the period of the cycle. We assume that the cyclic QHE has reached its time-periodic limit cycle at $t=0$, after a transient warming-up operation stage \cite{Liu.21.PRL}. 

For strong-coupling cyclic QHEs in the limit-cycle phase, it was recently emphasized that the first law of thermodynamics should take the following complete form \cite{Liu.21.PRL}
\begin{equation}\label{eq:first_law}
J_W+\sum_vJ_{Q_v}-J_A~=~0.
\end{equation}
Here, $J_{W,Q_v,A}$ are cycle-averaged thermodynamic fluxes \cite{Brandner.15.PRX,Brandner.16.PRE,Brandner.17.PRLa,Miller.21.PRL} corresponding to the work, heat and system-bath coupling ($\alpha=W,Q_v,A$)
\begin{equation}\label{eq:definitions}
J_{\alpha}~\equiv~ \frac{1}{\mathcal{T}}\int_0^{\mathcal{T}}dt\,d_t\langle \mathcal{O}_{\alpha}\rangle,
\end{equation}
where $\mathcal{O}_{W}=H(t)$, $\mathcal{O}_{Q_v}=-H_B^v$ and $\mathcal{O}_A=H_I(t)$, respectively; noting $\int_0^{\mathcal{T}}dt\,d_t\langle H_S(t)\rangle=0$ at limit cycle. $\langle \mathcal{O}\rangle\equiv \mathrm{Tr}[\rho(t)\mathcal{O}]$ denotes an ensemble average of any observable $\mathcal{O}$ over the global density matrix $\rho(t)$ of the composite system $H(t)$, $d_t\mathcal{O}\equiv d\mathcal{O}/dt$. In our convention, a heat engine mode corresponds to $J_W<0$, $J_{Q_h}>0$ and $J_{Q_c}<0$. We point out that $J_W$ encompasses work contributions from both driving the working medium and tuning on/off the interaction \cite{Wiedmann.20.NJP} since $d_t\langle H(t)\rangle=\langle d_t H(t)\rangle$ and $J_A\cdot\mathcal{T}$ accounts for the energy accumulated in the interaction term over a limit cycle. We focus on typical setups with $[H_I,H_S+H_B]\neq 0$ such that $J_A$ vanishes only at weak couplings in the limit-cycle phase \cite{Liu.21.PRL}. The mean entropy production rate $\sigma$ over a limit cycle is given by 
\begin{equation}\label{eq:entropy}
\sigma~=~-\sum_v\beta_vJ_Q^v,
\end{equation}
Here, $\beta_v=1/T_v$ are the inverse temperature of heat baths.

We consider a small temperature difference $\Delta T/T_v\ll 1$ with $\Delta T=T_h-T_c$, thereby allowing for a linear-response description of cyclic QHEs. Combining Eqs. (\ref{eq:first_law}) and (\ref{eq:entropy}), we find $\sigma=\beta_c(J_W-J_A)+(\beta_c-\beta_h)J_Q^h$ which motivates us to introduce thermodynamic affinities $\mathcal{F}_W=\beta_c$ and $\mathcal{F}_Q=\beta_c-\beta_h>0$ together with a renormalized work flux $J_{\widetilde{W}}\equiv J_W-J_A$ and a heat flux $J_Q\equiv J_{Q_h}$. We remark that by introducing a renormalized work flux we aim to develop a linear-response description that naturally incorporates as a special limit the existing version for cyclic QHEs at weak couplings (see, e.g., Refs \cite{Brandner.15.PRX,Brandner.16.PRE}) where $J_A$ vanishes.

Within linear irreversible thermodynamics \cite{Callen.85.NULL}, we can write down equations relating fluxes and affinities 
\bea\label{eq:linear_relation}
J_{\widetilde{W}} &=& L_{\widetilde{W}\widetilde{W}}\mathcal{F}_W+L_{\widetilde{W}Q}\mathcal{F}_Q,\nonumber\\
J_Q &=& L_{Q\widetilde{W}}\mathcal{F}_W+L_{QQ}\mathcal{F}_Q.
\eea
The kinetic coefficients $L_{\alpha\beta}$ ($\alpha,\beta=\widetilde{W},Q$) introduced above can be casted into the so-called Onsager matrix
\begin{equation}
\mathbb{L}~=~\left(\begin{array}{cc}
L_{\widetilde{W}\widetilde{W}} & L_{\widetilde{W}Q}\\
L_{Q\widetilde{W}} & L_{QQ}
\end{array}\right).
\end{equation}
We now find $\sigma=\sum_{\alpha,\beta}L_{\alpha\beta}\mathcal{F}_{\alpha}\mathcal{F}_{\beta}=\frac{\boldsymbol{\mathcal{F}}^T(\mathbb{L}+\mathbb{L}^T)\boldsymbol{\mathcal{F}}}{2}\equiv\boldsymbol{\mathcal{F}}^T\mathbb{L}^s\boldsymbol{\mathcal{F}}$ with $\boldsymbol{\mathcal{F}}=(\mathcal{F}_W,\mathcal{F}_Q)^T$ a $2\times 1$ vector and $\mathbb{L}^s\equiv(\mathbb{L}+\mathbb{L}^T)/2$ being the symmetric part of the matrix $\mathbb{L}$; the superscript `$T$' denotes the transpose. The non-negativity of $\sigma$ thus indicates that the symmetric part $\mathbb{L}^s$ must be positive semidefinite \cite{Brandner.16.PRE}, leading to the following constraints on kinetic coefficients
\bea
&&L_{\widetilde{W}\widetilde{W}}~\geqslant~0,~~~L_{QQ}~\geqslant~0,\label{eq:constraint_2}\nonumber\\
&&L_{\widetilde{W}\widetilde{W}}L_{QQ}-\frac{1}{4}\left(L_{\widetilde{W}Q}+L_{Q\widetilde{W}}\right)^2~\geqslant~0.\label{eq:constraint_1}
\eea

Though mathematically straightforward, the above linear-response description is not directly applicable for the characterization of the performance of strong-coupling cyclic QHEs, noting that only part of $J_{\widetilde{W}}$ corresponds to the actual work flux. We circumvent this issue by further adopting the following separations for kinetic coefficients as can be inferred from the form $J_{\widetilde{W}}=J_W-J_A$ and the linear nature of Eq. (\ref{eq:linear_relation}) \footnote{Alternatively, one can obtain the same separations by firstly writing down linear equations connecting $J_{W,Q,A}$ and $\mathcal{F}_{W,Q,A}$; $\sigma=\sum_{\alpha=W,Q,A}\mathcal{F}_{\alpha}J_{\alpha}$, and then combining the coefficients by noting $\mathcal{F}_A\equiv-\beta_c=-\mathcal{F}_W$.}: $L_{\widetilde{W}\widetilde{W}}=L_{WW}-L_{WA}+L_{AA}-L_{AW}$, $L_{\widetilde{W}Q}=L_{WQ}-L_{AQ}$ and $L_{Q\widetilde{W}}= L_{QW}-L_{QA}$, yielding
\bea\label{eq:jwq}
J_W &=& (L_{WW}-L_{WA})\mathcal{F}_W+L_{WQ}\mathcal{F}_Q,\nonumber\\
J_Q &=& (L_{QW}-L_{QA})\mathcal{F}_W+L_{QQ}\mathcal{F}_Q.
\eea
At weak couplings, we should have $L_{\alpha A}= L_{A\alpha}=0$ since $J_A=(L_{AW}-L_{AA})\mathcal{F}_W+L_{AQ}\mathcal{F}_Q=0$ where both $\mathcal{F}_W$ and $\mathcal{F}_Q$ are generally nonzero, reducing Eq. (\ref{eq:jwq}) to those for weak coupling scenarios \cite{Brandner.15.PRX,Brandner.16.PRE}. 

To facilitate an analytical treatment, one can introduce dimensionless parameters as combinations of kinetic coefficients \cite{Benenti.11.PRL}. For strong-coupling linear cyclic QHEs, we find the following four dimensionless parameters are adequate to describe thermodynamics and characterize optimal performance,
\bea\label{eq:parameter}
&&x~\equiv~\frac{L_{WQ}}{L_{QW}-L_{QA}},~~~y~\equiv~\frac{\mathcal{D}}{L_{QQ}(L_{WW}-L_{WA})-\mathcal{D}},\nonumber\\
&&z_1~\equiv~\frac{L_{WQ}}{L_{WQ}-L_{AQ}},~~~z_2~\equiv~\frac{\mathcal{D}}{L_{QQ}(L_{AA}-L_{AW})+\mathcal{D}}.
\eea
Here, $\mathcal{D}\equiv(L_{WQ}-L_{AQ})(L_{QW}-L_{QA})$. At weak couplings, expressions for $x$ and $y$ reduce to their well-adopted forms in systems with just work and heat fluxes \cite{Benenti.11.PRL,Brandner.15.PRX,Brandner.16.PRE} and $z_{1,2}$ become unity. The presence of two extra parameters $z_{1,2}\neq 1$ thus distinguishes strong-coupling cyclic QHEs from weak-coupling counterparts in the linear response regime. The ratio $x/z_1$ characterizes the degree of time-reversal symmetry breaking at strong couplings, in analog with the weak coupling scenario \cite{Brandner.15.PRX,Brandner.16.PRE}. 

\begin{figure}[thb!]
 \centering
\includegraphics[width=0.9\columnwidth]{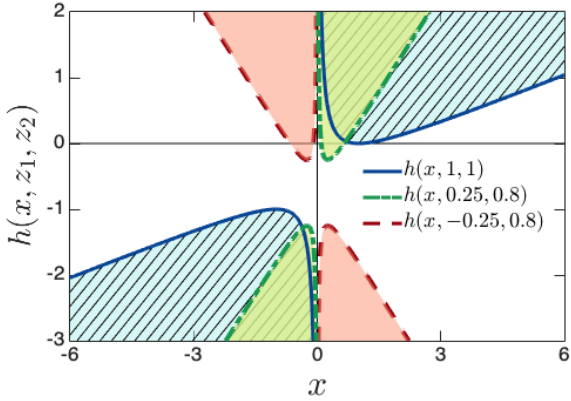} 
%\vspace*{-16mm}
\caption{Bound $h(x,z_1,z_2)$ [cf. Eq. (\ref{eq:yy})] as a function of $x$ with $z_{1,2}=1$ (blue solid line, corresponding to the weak coupling limit), $z_1=0.25$, $z_2=0.8$ (green dashed-dotted line) and $z_1=-0.25$, $z_2=0.8$ (red dashed line). $y^{-1}$ can take values only from the shaded and hatched regions enclosed by $h(x,z_1,z_2)$.
}
\protect\label{fig:h_bound}
\end{figure}

In terms of $x,y,z_1,z_2$ and using separations of kinetic coefficients introduced above, the constraints in Eq. (\ref{eq:constraint_1}) transfer to
\begin{equation}\label{eq:inequality_1}
\frac{1}{xz_1}\left(\frac{1}{y}+\frac{1}{z_2}\right)-\frac{1}{4}\left(\frac{1}{x}+\frac{1}{z_1}\right)^2~\geqslant~0,
\end{equation}
which yields
\bea\label{eq:yy}
y^{-1} &\geqslant& h(x,z_1,z_2),~~~\mathrm{for}~xz_1\geqslant 0,\nonumber\\
y^{-1} &\leqslant& h(x,z_1,z_2),~~~\mathrm{for}~xz_1< 0.
\eea
Here, we have introduced $h(x,z_1,z_2)\equiv [z_2(x+z_1)^2-4xz_1]/(4xz_1z_2)$. At weak couplings with $z_{1,2}=1$, the above inequalities reduce to known constraints on $y$: $0\leqslant y\leqslant 4x/(x-1)^2$ ($4x/(x-1)^2\leqslant y\leqslant 0$) for $x\geqslant 0$ ($x<0$) \cite{Benenti.11.PRL,Brandner.15.PRX,Brandner.16.PRE}. It can be verified that $h(x,z_1,z_2)\geqslant(z_2-1)/z_2$ ($\leqslant-1/z_2$) in the region of $xz_1\geqslant0$ ($xz_1< 0$). Furthermore, we note that $h(-x,-z_1,z_2)=h(x,z_1,z_2)$ which leaves the bounds in Eq. (\ref{eq:yy}) unchanged. In Fig. \ref{fig:h_bound}, we depict a set of results for $h(x,z_1,z_2)$ with varying $z_{1,2}$ which clearly verifies the aforementioned features of $h(x,z_1,z_2)$. By contrasting the blue hatched and green shaded regions depicted in Fig. \ref{fig:h_bound}, one can observe that the allowed parameter region of $y^{-1}$ shrinks and moves downwards when $z_{1,2}$ deviate from the weak coupling limit.   
%
%-------------------------------------------------
%-------------------------------------------------
Particularly, $y$ can take negative values in the region of $xz_1\geqslant0$ when $z_{1,2}\neq 1$, in direct contrast to the weak coupling limit where $y$ is non-negative in the region of $x\geqslant 0$. As will be seen later, the changes in $h(x,z_1,z_2)$ lead to profound consequences on thermodynamic bounds on optimal efficiencies of linear cyclic QHEs. 

{\it Optimized performance.--}  
Using Eq. (\ref{eq:jwq}), the output power $P\equiv-T_c\mathcal{F}_WJ_W$ and thermodynamic efficiency $\eta\equiv P/J_Q$ in the linear response regime are given by
\bea
P &=& -T_c\mathcal{F}_W[(L_{WW}-L_{WA})\mathcal{F}_W+L_{WQ}\mathcal{F}_Q],\label{eq:PP}\\
\eta &=& -\frac{T_c\mathcal{F}_W[(L_{WW}-L_{WA})\mathcal{F}_W+L_{WQ}\mathcal{F}_Q]}{(L_{QW}-L_{QA})\mathcal{F}_W+L_{QQ}\mathcal{F}_Q}.\label{eq:eta}
\eea
We consider efficiency at maximum power (EMP) and maximum efficiency (ME) as figures of merit characterizing the optimal performance of linear cyclic QHEs from weak to strong couplings. Particularly, we are interested in general thermodynamic bounds on both the EMP and ME. To ensure the existence of {\it non-negative} EMP and ME for linear cyclic QHEs with $y$ satisfying Eq. (\ref{eq:yy}), we find that one should limit the ranges of $z_{1,2}$ to (see details in Supplemental Material \cite{SM})
\begin{equation}\label{eq:z_range}
0\leqslant z_1\leqslant 1,~~~~\frac{1}{2-z_1}\leqslant z_2\leqslant 1.
\end{equation}
We require that one can take $z_{2}=1$ only when $z_{1}=1$ and vice versa. Eq. (\ref{eq:z_range}) is a direct result of the non-negativity of both the entropy production rate and optimal efficiencies, no extra assumptions are invoked besides limiting $z_1$ to a positive number due to $h(-x,-z_1,z_2)=h(x,z_1,z_2)$. In \cite{SM}, we address the scenario with a negative $z_1$ and show that the results and conclusions obtained below remain unaltered.

We first focus on the EMP and its thermodynamic bound. By maximizing the output power [cf. Eq. (\ref{eq:PP})] with respect to $\mathcal{F}_W$ \footnote{This optimization corresponds to varying the temperatures of thermal baths while fixing the temperature difference.}, we receive an optimal condition $\mathcal{F}_W^o=-L_{WQ}\mathcal{F}_Q/[2(L_{WW}-L_{WA})]$. Then we can obtain the EMP $\eta(P_{\mathrm{max}})=\left.P/J_Q\right|_{\mathcal{F}_W=\mathcal{F}_W^o}$ as
\begin{equation}\label{eq:emp}
\eta(P_{\mathrm{max}}) 
~=~ \frac{\eta_{c}}{2}\frac{xyz_1}{2(1+y)-yz_1}.
\end{equation}
Here, $\eta_c=1-T_c/T_h$ denotes the Carnot limit. In arriving at the above equation, we have used the replacement $[L_{QQ}(L_{WW}-L_{WA})]/L_{WQ}^2=\left(y^{-1}+1\right)/(xz_1)$. When $z_1=1$, we recover the known expression for $\eta(P_{\mathrm{max}})$\cite{Benenti.11.PRL,Brandner.15.PRX}.

Since $\eta(P_{\mathrm{max}})$ is a decreasing (an increasing) function of $y^{-1}$ when $xz_1\geqslant 0$ ($xz_1<0$), $\eta(P_{\mathrm{max}})$ attains its maximum when $y^{-1}=h(x,z_1,z_2)$, yielding a thermodynamic upper bound $\eta_{\mathrm{EMP}}$ on the EMP ($\eta(P_{\mathrm{max}})\leqslant \eta_{\mathrm{EMP}}$)
\begin{equation}\label{eq:eta_P_max}
\eta_{\mathrm{EMP}}(x,z_{1,2})\equiv\frac{x^2z_1^2z_2\eta_{c}}{z_2(x+z_1)^2+2xz_1(2z_2-2-z_1z_2)}.
\end{equation}
At weak couplings where $z_{1,2}=1$, $\eta_{\mathrm{EMP}}^{\mathrm{weak}}(x)=\eta_cx^2/(x^2+1)$ which is the known bound on EMP obtained previously \cite{Benenti.11.PRL,Brandner.15.PRX}. Away from the weak coupling limit, $\eta_{\mathrm{EMP}}$ is in general not a symmetric function of $x$. From Eq. (\ref{eq:eta_P_max}), we can deduce the following properties of $\eta_{\mathrm{EMP}}$ 
\begin{equation}\label{eq:eta_lim}
\eta^{\infty}_{\mathrm{EMP}}~=~ z_1^2\eta_c,~~\mathrm{and}~~
\eta_{\mathrm{EMP}}(x/z_1,z_{1,2})~\leqslant~ \eta_{\mathrm{EMP}}^{\mathrm{weak}}(x/z_1).
\end{equation}
Here, $\eta^{\infty}_{\mathrm{EMP}}~\equiv~\lim\limits_{|x|\to\infty}\eta_{\mathrm{EMP}}$; noting $|x|\to\infty$ corresponds to a rather strong time-reversal symmetry breaking since $z_1$ is finite. The proof of the inequality in Eq. (\ref{eq:eta_lim}) can be found in \cite{SM}. We remark that here we compared the $\eta_{\mathrm{EMP}}$ with the same degree of time-reversal symmetry breaking $x/z_1$ which enables a fair comparison between weak-coupling and strong-coupling linear cyclic QHEs; similar for Eq. (\ref{eq:eta_m_lim}) below. Eq. (\ref{eq:eta_lim}) represents our first main result. 
%
%-------------------------------------------------
\begin{figure}[thb!]
 \centering
\includegraphics[width=1\columnwidth]{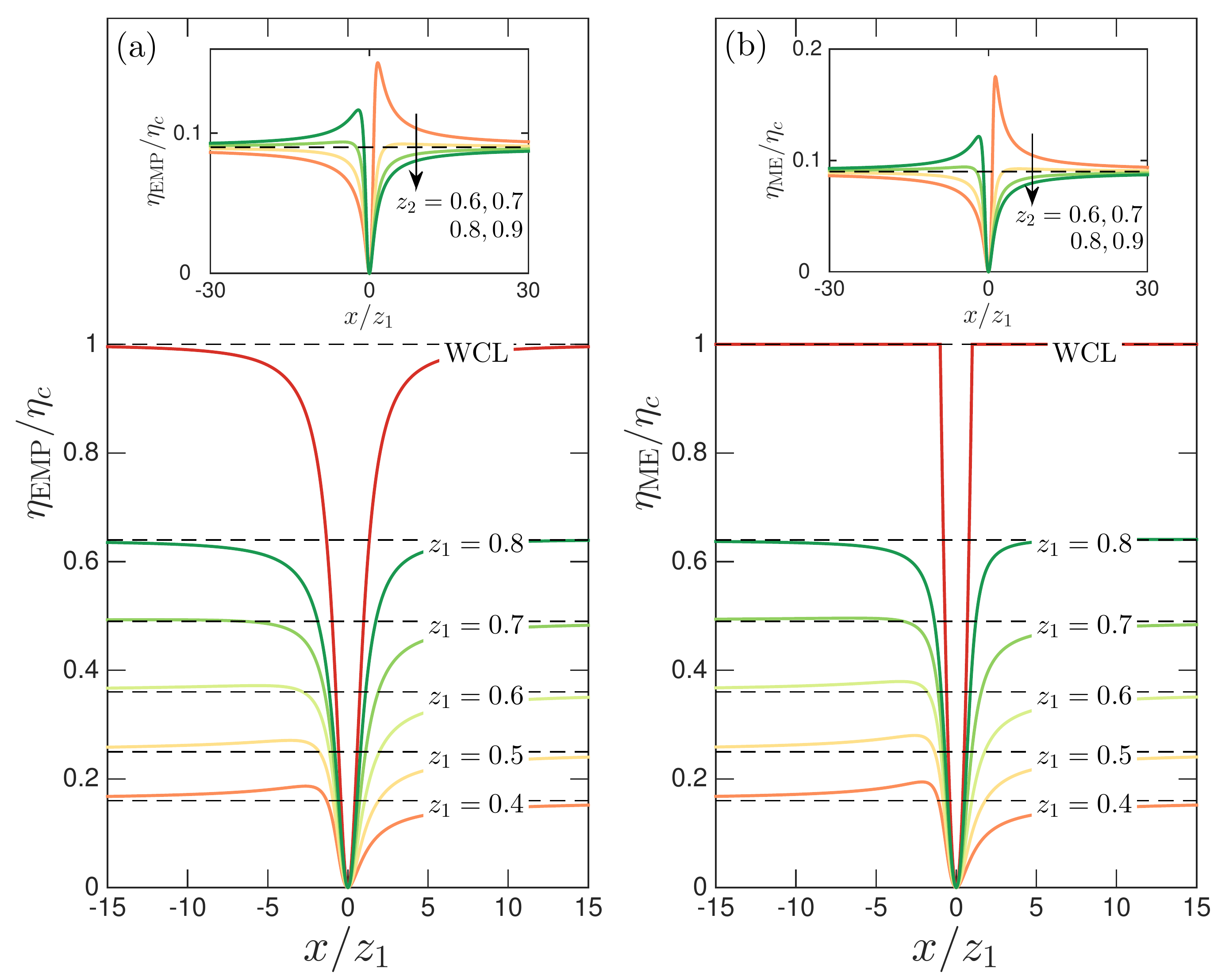} 
%\vspace*{-16mm}
\caption{(a) $\eta_{\mathrm{EMP}}/\eta_c$ [cf. Eq. (\ref{eq:eta_P_max})] as a function of $x/z_1$ with varying $z_1$ and fixed $z_2=0.9$. Inset: Results with varying $z_2$ and fixed $z_1=0.3$. (b) $\eta_{\mathrm{ME}}/\eta_c$ [cf. Eq. (\ref{eq:eta_max})] as a function of $x/z_1$ with varying $z_1$ and fixed $z_2=0.9$. Inset: Results with varying $z_2$ and fixed $z_1=0.3$. For comparisons, we depict the weak coupling limit (WCL) with $z_{1,2}=1$. Dashed horizontal lines in both plots mark the value of $z_1^2$.
}
\protect\label{fig:optimal}
\end{figure}
%-------------------------------------------------
A typical set of results for $\eta_{\mathrm{EMP}}$ as a function of $x/z_1$ with varying $z_{1,2}$ satisfying Eq. (\ref{eq:z_range}) is depicted in Fig. \ref{fig:optimal} (a). From Fig. \ref{fig:optimal} (a), it is apparent that $\eta_{\mathrm{EMP}}$ remains smaller than its weak-coupling limit with $z_{1,2}=1$, and saturates $z_1^2\eta_c$ when $|x|\to\infty$.

We then turn to the ME and its thermodynamic bound. To get the ME $\eta_{\mathrm{max}}$, we directly optimize Eq. (\ref{eq:eta}) with respect to $\mathcal{F}_W$. After some lengthy algebra, we find (see details in \cite{SM})
\begin{equation}\label{eq:eta_m_p}
\eta_{\mathrm{max}} ~=~\eta_c\frac{|x|}{|z_1y|}\left(\sqrt{|1+y|}-\sqrt{|1+y-z_1y|}\right)^2
\end{equation}
Here, $|\mathcal{O}|$ takes the absolute value of $\mathcal{O}$. When $z_1=1$, we get $\eta_{\mathrm{max}}=\eta_c \frac{x}{y}(\sqrt{y+1}-1)^2=\eta_c x\frac{\sqrt{y+1}-1}{\sqrt{y+1}+1}$ by noting $y\geqslant -1$ and $x,~y$ have the same sign, recovering the expression for $\eta_{\mathrm{max}}$ used in the weak coupling limit \cite{Benenti.11.PRL,Brandner.15.PRX,Brandner.16.PRE}. It can be easily verified that $\eta_{\mathrm{max}}$ is a decreasing (an increasing) function of $y^{-1}$ when $xz_1\geqslant 0$ ($xz_1<0$). Hence, similar to the EMP, we can obtain a thermodynamic upper bound $\eta_{\mathrm{ME}}$ on the ME by taking $y^{-1}=h(x,z_1,z_2)$ in Eq. (\ref{eq:eta_m_p}) ($\eta_{\mathrm{max}}\leqslant \eta_{\mathrm{ME}}$):
\begin{equation}\label{eq:eta_max}
\eta_{\mathrm{ME}}(x,z_{1,2})~\equiv~ \eta_c|x|\left(\sqrt{\left|\frac{1}{z_1}(h+1)\right|}-\sqrt{\left|\frac{1}{z_1}(h+1)-1\right|}\right)^2,
\end{equation}
which again reduces to the known bound $\eta_{\mathrm{ME}}^{\mathrm{weak}}(x)=\eta_c x^2$ ($\eta_c$) for $|x|\leqslant 1$ ($|x|\geqslant 1$)\cite{Benenti.11.PRL} when $z_{1,2}=1$. From the above bound, we find that
\begin{equation}\label{eq:eta_m_lim}
\eta_{\mathrm{ME}}^{\infty}~=~z_1^2\eta_c,~~\mathrm{and}~~\eta_{\mathrm{ME}}(x/z_1,z_{1,2})~\leqslant~\eta_{\mathrm{ME}}^{\mathrm{weak}}(x/z_1).
\end{equation}
Here, $\eta_{\mathrm{ME}}^{\infty}~\equiv~\lim\limits_{|x|\to\infty}\eta_\mathrm{ME}$. The proof of the inequality in Eq. (\ref{eq:eta_m_lim}) can be found in \cite{SM}. Eq. (\ref{eq:eta_m_lim}) is our second main result. A typical set of results for $\eta_\mathrm{ME}$ as a function of $x/z_1$ with varying $z_{1,2}$ satisfying Eq. (\ref{eq:z_range}) is presented in Fig. \ref{fig:optimal} (b) which clearly validates the properties listed in Eq. (\ref{eq:eta_m_lim}).

Combining Eqs. (\ref{eq:eta_lim}) and (\ref{eq:eta_m_lim}), we can draw the following general conclusions concerning the role of system-bath coupling in shaping the optimal performance of linear cyclic QHEs ($n=\mathrm{EMP}, \mathrm{ME}$): Most significantly, (i) $\eta_n$ is upper bounded by its weak coupling limit, implying that an {\it optimal} weak-coupling cyclic QHE performs more efficiently than its strong-coupling counterpart in the linear response regime. (ii) Noting the fact in (i) and that $\eta_n$ of weak-coupling linear cyclic QHEs attains its maximum $\eta_c$ under strong time-reversal symmetry breaking as $|x|\to\infty$, one can infer that optimal linear cyclic QHEs can reach the Carnot limit $\eta_c$ only in the weak coupling limit. (iii) Away from the weak-coupling limit as $z_1$ decreases from 1, the extreme value $\eta_n^{\infty}=z_1^2\eta_c$ of $\eta_n$ drops quadratically in $z_1$ relative to the Carnot limit. We can further relate $z_1$ to the dimensionless system-bath coupling strength $\lambda$: Denoting $H_I=\lambda \tilde{H}_I$ with $\tilde{H}_I$ a rescaled system-bath interaction, we have $J_A\propto\lambda$ by noting the definition Eq. (\ref{eq:definitions}) and hence $L_{\alpha A},L_{A\alpha}\propto\lambda$ since the affinities are $\lambda$-independent, leading to $z_1\simeq1-c_1\lambda+\mathcal{O}(\lambda^2)$ with $c_1$ a model-dependent coefficient (noting definition in Eq. (\ref{eq:parameter})). Therefore, we expect a relative suppression $(\eta^{\infty,\mathrm{weak}}_n-\eta^{\infty}_n)/\eta^{\infty,\mathrm{weak}}_n=1-z_1^2\propto \lambda+\mathcal{O}(\lambda^2)$ scales at least linearly in $\lambda$ with $\eta^{\infty,\mathrm{weak}}_n=\eta_c$ for weak-coupling linear cyclic QHEs. We emphasize that the aforementioned conclusions hold regardless of details of cyclic QHEs (i.e., the detailed form of $H(t)$ in Eq. (\ref{eq:HH})) provided the temperature difference between the baths is small.

{\it Discussion.--}It is interesting to explore whether optimal weak-coupling cyclic QHEs can outperform their strong-coupling counterparts beyond the linear response regime. To provide a hint, consider a reversible thermodynamic cycle with vanishing entropy production which usually necessitates the ME. Specifically, we have $S=-\sum_v\beta_vQ_v=\beta_c(W-A)+(\beta_c-\beta_h)Q_h=0$ ($S=\mathcal{T}\sigma$, $Q_v=\mathcal{T}J_{Q_v}$, $W=\mathcal{T}J_W$ and $A=\mathcal{T}J_A$), yielding $W=A-\eta_cQ_h$ with $\eta_c$ the Carnot efficiency. Inserting the expression for $W$ into the definition of efficiency $\eta=-W/Q_h$, we get the ME $\eta_{\mathrm{max}}=\eta_c-A/Q_h$. Recognizing that optimal strong-coupling cyclic QHEs with normal thermal baths cannot break the Carnot limit, we should have $A>0$ at strong couplings by noting $Q_h>0$; $A$ remains nonzero at strong couplings as long as $W\neq 0$ \cite{Liu.21.PRL}, consequently, one naturally infers $\eta_{\mathrm{max}}\leqslant \eta_{\mathrm{max}}^{\mathrm{weak}}=\eta_c$. Hence, we conjecture that strong coupling will likely suppress the ME of cyclic QHEs beyond the linear response regime. We leave possible validations to future works.

To correctly interpret the present results, it is necessary to discriminate between optimal and non-optimal QHEs. Taking a set of $z_{1,2}<1$, we only stated that $\eta_n(z_{1,2})<\eta_n^{\mathrm{weak}}\equiv\eta_n(z_{1,2}=1)$ with $n=\mathrm{EMP}, \mathrm{ME}$ in the linear response regime. However, if one just considers a non-optimal linear QHE with an actual efficiency $\eta<\eta_n$, it is possible to have the trend $\eta(z_{1,2})>\eta^{\mathrm{weak}}\equiv\eta(z_{1,2}=1)$ as opposed to the relative relation for the optimal efficiency $\eta_n$, namely, our present results do not rule out the possibility of having non-optimal cyclic QHEs capable of benefitting from strong couplings in the linear response regime.

In summary, we analyzed the optimal performance of generic cyclic QHEs from weak to strong couplings in the linear response regime and obtained thermodynamic bounds on optimal efficiencies. We revealed a universal feature of linear cyclic QHEs that system-bath coupling tends to suppress both the efficiency at maximum power and maximum efficiency. Under strong time-reversal symmetry breaking, this suppression scales at least linearly in system-bath coupling strength. Our results provide new insights into the investigation of the effects of system-bath coupling on heat-work conversion and are relevant for the search of efficient strong-coupling QHEs. 

We thank Dvira Segal for helpful discussions. J.L. acknowledges support from the startup funding of Shanghai University. K.A.J acknowledges support from the National Science Foundation Grant No. CHE-2154291.

%\bibliography{LRT}

%

\newpage
\renewcommand{\thesection}{\Roman{section}} 
\renewcommand{\thesubsection}{\Alph{subsection}}
\renewcommand{\theequation}{S\arabic{equation}}
\renewcommand{\thefigure}{S\arabic{figure}}
\renewcommand{\thetable}{S\arabic{table}}
\setcounter{equation}{0}  % reset counter
\setcounter{figure}{0}

\begin{widetext}

{\Large{\bf Supplemental material:} Optimal linear cyclic quantum heat engines cannot benefit from strong coupling}
\\
\\
\\

In this supplemental material, we first present derivation details regarding the maximum efficiency in Sec. I. Then, in Sec. II we identify the range of parameters $z_{1,2}$ which ensure the existence of non-negative efficiency at maximum power and maximum efficiency by limiting to a positive $z_1$. Next, in Sec. III we prove that both the efficiency at maximum power and maximum efficiency of cyclic quantum heat engines (QHEs) are upper bounded by their weak-coupling limits. Lastly, in Sec. IV, we address the case by allowing for a negative $z_1$ and revisit the results obtained previously.

\section{I. Deriving maximum efficiency}\label{sec:1}
To get the maximum efficiency $\eta_{\mathrm{max}}$, we optimize the following equation with respect to $\mathcal{F}_W$
\begin{equation}
\eta ~=~ -\frac{T_c\mathcal{F}_W[(L_{WW}-L_{WA})\mathcal{F}_W+L_{WQ}\mathcal{F}_Q]}{(L_{QW}-L_{QA})\mathcal{F}_W+L_{QQ}\mathcal{F}_Q},
\end{equation}
receiving a quadratic equation for $\mathcal{F}_W$:
\begin{equation}\label{eq:quadratic}
(L_{WW}-L_{WA})(L_{QW}-L_{QA})\mathcal{F}_W^2+2L_{QQ}(L_{WW}-L_{WA})\mathcal{F}_Q\mathcal{F}_W+L_{WQ}L_{QQ}\mathcal{F}_Q^2~=~0,
\end{equation}
from which we can get two solutions $\mathcal{F}_W^{\pm}$
\begin{equation}\label{eq:F_solution}
\mathcal{F}_W^{\pm}~=~\frac{1}{(L_{WW}-L_{WA})(L_{QW}-L_{QA})}\left[-L_{QQ}\mathcal{F}_Q(L_{WW}-L_{WA})\pm\sqrt{\Omega/4}\right].
\end{equation}
To ensure real values of $\mathcal{F}_W^{\pm}$, we should require
\bea\label{eq:Omega}
\Omega &\equiv& 4\mathcal{F}_Q^2L_{QQ}\left[L_{QQ}(L_{WW}-L_{WA})^2-L_{WQ}(L_{WW}-L_{WA})(L_{QW}-L_{QA})\right]\nonumber\\
&=& \frac{4\mathcal{F}_Q^2L_{WQ}^4}{x^2}\frac{1}{z_1}\left(\frac{1}{y}+1\right)\left[\frac{1}{z_1}\left(\frac{1}{y}+1\right)-1\right]~\geqslant~0.
\eea
In deriving the second line, we have utilized the relation by noting the definitions for $x,y,z_1$
\begin{equation}\label{eq:auxi}
\frac{L_{QQ}(L_{WW}-L_{WA})}{L_{WQ}^2}~=~\frac{1}{xz_1}\left(\frac{1}{y}+1\right).
\end{equation} 
At the moment, we take $\Omega\geqslant 0$. In Sec. II below, we will explicitly derive the conditions which ensure a non-negative $\Omega$. Inserting the above solutions in Eq. (\ref{eq:F_solution}) into the definition $J_Q= (L_{QW}-L_{QA})\mathcal{F}_W+L_{QQ}\mathcal{F}_Q$, we find
\begin{equation}\label{eq:JQ_pm}
J_Q^{\pm}~=~\pm\frac{\sqrt{\Omega}}{2(L_{WW}-L_{WA})}~=~\pm\frac{\mathcal{F}_QL_{WQ}^2\sqrt{\Theta}}{(L_{WW}-L_{WA})|x|}
\end{equation}
corresponding to solution $\mathcal{F}_W^{\pm}$, respectively. Here, $|\mathcal{O}|$ takes the absolute value of $\mathcal{O}$, we have denoted $\Theta\equiv\frac{1}{z_1}\left(\frac{1}{y}+1\right)\left[\frac{1}{z_1}\left(\frac{1}{y}+1\right)-1\right]=\left|\frac{1}{z_1}\left(\frac{1}{y}+1\right)\right|\cdot\left|\frac{1}{z_1}\left(\frac{1}{y}+1\right)-1\right|$ by noting that $\frac{1}{z_1}\left(\frac{1}{y}+1\right)$ and $\frac{1}{z_1}\left(\frac{1}{y}+1\right)-1$ have the same sign according to Eq. (\ref{eq:Omega}). 

For the case with $L_{WW}-L_{WA}\geqslant 0$ (it implies $L_{QQ}(L_{WW}-L_{WA})/L_{WQ}^2\geqslant 0$ [cf. Eq. (\ref{eq:auxi})]), we should pick the solution $\mathcal{F}_W^+$ in Eq. (\ref{eq:F_solution}) and find
\bea
\eta_{\mathrm{max}}^+ &=& \frac{\left.P\right|_{\mathcal{F}_W=\mathcal{F}_W^+}}{J_Q^+}\nonumber\\
&=& T_c\mathcal{F}_Q\frac{|x|}{\sqrt{\Theta}}\Bigg\{-\frac{L_{QQ}^2(L_{WW}-L_{WA})^2}{L_{WQ}^4}x^2-\Theta+x\left(2\frac{x}{|x|}\sqrt{\Theta}+1\right)\frac{L_{QQ}(L_{WW}-L_{WA})}{L_{WQ}^2}-\frac{x}{|x|}\sqrt{\Theta}\Bigg\}\nonumber\\
&=& \eta_c\frac{|x|}{\sqrt{\Theta}}\Bigg\{-\left[\frac{1}{z_1}\left(\frac{1}{y}+1\right)\right]^2-\Theta+\left(2\frac{x}{|x|}\sqrt{\Theta}+1\right)\frac{1}{z_1}\left(\frac{1}{y}+1\right)-\frac{x}{|x|}\sqrt{\Theta}\Bigg\}.
%&=& T_c\mathcal{F}_Q x\left[-2\frac{|x|}{x}\sqrt{\Theta}+2\frac{1}{z_1}\left(\frac{1}{y}+1\right)-1\right]\nonumber\\
%&=& T_c\mathcal{F}_Q x \frac{(\sqrt{y+1}-\sqrt{y+1-z_1y})^2}{yz_1}\nonumber\\
%&=& \eta_c x \frac{\sqrt{y+1}-\sqrt{y+1-z_1y}}{\sqrt{y+1}+\sqrt{y+1-z_1y}}.
\eea
In deriving the last line, we utilized Eq. (\ref{eq:auxi}). Noting the presence of $|x|$, it is convenient to treat $\eta_{\mathrm{max}}^+(x\geqslant 0)$ and $\eta_{\mathrm{max}}^+(x<0)$ separately. For $x\geqslant 0$, $L_{QQ}(L_{WW}-L_{WA})/L_{WQ}^2\geqslant 0$ [cf. Eq. (\ref{eq:auxi})] together with $\Omega \geqslant 0$ [cf. Eq. (\ref{eq:Omega})] indicate that we should take $\frac{1}{z_1}(1/y+1)\geqslant 1$, consequently, both $\frac{1}{z_1}(1/y+1)$ and $\frac{1}{z_1}(1/y+1)-1$ are positive, hence we have
\bea\label{eq:pp}
\eta_{\mathrm{max}}^+(x\geqslant 0) &=& \eta_c\frac{x}{\sqrt{\Theta}}\left[-2\Theta+2\sqrt{\Theta}\frac{1}{z_1}\left(
\frac{1}{y}+1\right)-\sqrt{\Theta}\right]\nonumber\\
&=& \eta_c x\left(\sqrt{\frac{y+1}{z_1y}}-\sqrt{\frac{y+1-z_1y}{z_1y}}\right)^2~=~\eta_c x\left(\sqrt{\left|\frac{y+1}{z_1y}\right|}-\sqrt{\left|\frac{y+1-z_1y}{z_1y}\right|}\right)^2\nonumber\\
&=& \eta_c\frac{x}{|z_1y|}\left(\sqrt{|y+1|}-\sqrt{|y+1-z_1y|}\right)^2.
\eea
While for $x<0$, we should instead take $\frac{1}{z_1}(1/y+1)<0$ in order to satisfy both $L_{QQ}(L_{WW}-L_{WA})/L_{WQ}^2\geqslant 0$ [cf. Eq. (\ref{eq:auxi})] and $\Omega \geqslant 0$ [cf. Eq. (\ref{eq:Omega})]. Taking into account the fact that both $\frac{1}{z_1}(1/y+1)$ and $\frac{1}{z_1}(1/y+1)-1$ are negative in this case, we find
\bea\label{eq:mm}
\eta_{\mathrm{max}}^+(x<0) &=& -\eta_c\frac{x}{\sqrt{\Theta}}\left[-2\Theta-2\sqrt{\Theta}\frac{1}{z_1}\left(
\frac{1}{y}+1\right)+\sqrt{\Theta}\right]\nonumber\\
&=& -\eta_c x\left[-2\sqrt{\Theta}-2\frac{1}{z_1}\left(\frac{1}{y}+1\right)+1\right]~=~ -\eta_c x\left[-2\sqrt{\Theta}+\left|2\frac{1}{z_1}\left(\frac{1}{y}-1\right)+1\right|\right]\nonumber\\
&=& -\eta_c x\left(\sqrt{\left|\frac{y+1}{z_1y}\right|}-\sqrt{\left|\frac{y+1-z_1y}{z_1y}\right|}\right)^2\nonumber\\
&=& -\eta_c\frac{x}{|z_1y|}\left(\sqrt{|y+1|}-\sqrt{|y+1-z_1y|}\right)^2.
\eea
In arriving at the second line, we noticed that $2\frac{1}{z_1}\left(\frac{1}{y}+1\right)-1$ is always negative. In third line, we used the form $\Theta=\left|\frac{1}{z_1}\left(\frac{1}{y}+1\right)\right|\cdot\left|\frac{1}{z_1}\left(\frac{1}{y}+1\right)-1\right|$ (see definition below Eq. (\ref{eq:JQ_pm})). Combining Eqs. (\ref{eq:pp}) and (\ref{eq:mm}), we finally get
\begin{equation}
\eta_{\mathrm{max}}^+ ~=~ \eta_c\frac{|x|}{|z_1y|}\left(\sqrt{|y+1|}-\sqrt{|y+1-z_1y|}\right)^2
\end{equation}
which is just Eq. (18) in the main text. It is evident that $\eta_{\mathrm{max}}^+$ is always positive.

To complete the derivations, we show in the following that the solution $\mathcal{F}_W^-$ always yields a negative maximum efficiency which is clearly unphysical. To have $J_Q^->0$, we should let $L_{WW}-L_{WA}<0$ which then implies $L_{QQ}(L_{WW}-L_{WA})/L_{WQ}^2<0$ [cf. Eq. (\ref{eq:auxi})]. We find
\bea
\eta_{\mathrm{max}}^- &=& \frac{\left.P\right|_{\mathcal{F}_W=\mathcal{F}_W^-}}{J_Q^-}\nonumber\\
&=& T_c\mathcal{F}_Q\frac{|x|}{\sqrt{\Theta}}\Bigg\{\frac{L_{QQ}^2(L_{WW}-L_{WA})^2}{L_{WQ}^4}x^2+\Theta+x\left(2\frac{x}{|x|}\sqrt{\Theta}-1\right)\frac{L_{QQ}(L_{WW}-L_{WA})}{L_{WQ}^2}-\frac{x}{|x|}\sqrt{\Theta}\Bigg\}\nonumber\\
&=& \eta_c |x| \left[2\sqrt{\Theta}+2\frac{x}{|x|}\frac{1}{z_1}\left(\frac{1}{y}+1\right)-\frac{x}{|x|}\right].
\eea
In getting the last line, we have utilized Eq. (\ref{eq:auxi}). Starting from the above expression, we can have
\bea\label{eq:d6}
\eta_{\mathrm{max}}^-(x\geqslant 0) &=& \eta_c x\left[2\sqrt{\Theta}+2\frac{1}{z_1}\left(\frac{1}{y}+1\right)-1\right]\nonumber\\
&=& \eta_c x\left[2\sqrt{\Theta}-\left|2\frac{1}{z_1}\left(\frac{1}{y}+1\right)-1\right|\right]\nonumber\\
&=& -\eta_c x\left(\sqrt{\left|\frac{y+1}{z_1y}\right|}-\sqrt{\left|\frac{y+1-z_1y}{z_1y}\right|}\right)^2<0.
\eea
Here, we noted that $2\frac{1}{z_1}\left(\frac{1}{y}+1\right)-1$ is always negative since $L_{QQ}(L_{WW}-L_{WA})/L_{WQ}^2<0$, $x\geqslant 0$ and $\Omega\geqslant 0$ [cf. Eq. (\ref{eq:Omega})]. Similarly,
\bea\label{eq:d7}
\eta_{\mathrm{max}}^-(x<0) &=& -\eta_c x \left[2\sqrt{\Theta}-2\frac{1}{z_1}\left(\frac{1}{y}+1\right)+1\right]\nonumber\\
&=& \eta_c x\left[-2\sqrt{\Theta}+2\frac{1}{z_1}\left(\frac{1}{y}+1\right)-1\right]\nonumber\\
&=& \eta_c x\left(\sqrt{\frac{y+1}{z_1y}}-\sqrt{\frac{y+1-z_1y}{z_1y}}\right)^2<0.
\eea
Here, we noted that $2\frac{1}{z_1}\left(\frac{1}{y}+1\right)-1$ is always positive since $L_{QQ}(L_{WW}-L_{WA})/L_{WQ}^2<0$, $x< 0$ and $\Omega\geqslant 0$ [cf. Eq. (\ref{eq:Omega})]. Eqs. (\ref{eq:d6}) and (\ref{eq:d7}) imply that $\eta_{\mathrm{max}}^-$ is always negative even though $J_Q^->0$, hence $\mathcal{F}_W^-$ in Eq. (\ref{eq:F_solution}) is not a physical solution.

\section{II. Constraints on the ranges of $z_{1,2}$}\label{sec:2}
\subsection{A. Conditions for non-negative efficiency at maximum power}\label{subsec:1}
To enable a non-negative efficiency at maximum power at strong couplings, it is enough to require $P_{\mathrm{max}}\geqslant 0$ and $\left.J_Q\right|_{\mathcal{F}_W=\mathcal{F}_W^o}\geqslant 0$; Recalling that $J_Q= (L_{QW}-L_{QA})\mathcal{F}_W+L_{QQ}\mathcal{F}_Q$ and $\mathcal{F}_W^o=-L_{WQ}\mathcal{F}_Q/[2(L_{WW}-L_{WA})]$ which maximizes the output power. Firstly, we have 
\bea\label{eq:a1}
&& P_{\mathrm{max}}=\frac{T_cL_{WQ}^2\mathcal{F}_Q^2}{4(L_{WW}-L_{WA})}\geqslant 0~\Rightarrow~L_{WW}-L_{WA}\geqslant 0\nonumber\\
&& \Rightarrow \frac{1}{xz_1}\left(\frac{1}{y}+1\right)\geqslant 0~\Rightarrow~\left\{\begin{array}{c}
y^{-1}\geqslant -1,~~\mathrm{for}~xz_1\geqslant 0,\\
y^{-1}\leqslant -1,~~\mathrm{for}~xz_1< 0.
\end{array}\right.
\eea
and
\bea\label{eq:a2}
&& \left.J_Q\right|_{\mathcal{F}_W=\mathcal{F}_W^o}~=~-\frac{(L_{QW}-L_{QA})L_{WQ}\mathcal{F}_Q}{2(L_{WW}-L_{WA})}+L_{QQ}\mathcal{F}_Q\geqslant 0\nonumber\\
&& \Rightarrow 2\frac{L_{QQ}(L_{WW}-L_{WA})}{L_{WQ}^2}\geqslant \frac{L_{QW}-L_{QA}}{L_{WQ}}~\Rightarrow~\frac{2}{xz_1}\left(\frac{1}{y}+1\right)\geqslant \frac{1}{x}\nonumber\\
&& \Rightarrow~\left\{\begin{array}{c}
y^{-1}\geqslant \frac{z_1}{2}-1,~~\mathrm{for}~xz_1\geqslant 0,\\
y^{-1}\leqslant -1,~~\mathrm{for}~xz_1< 0.
\end{array}\right.
\eea
Here, we have used Eq. (\ref{eq:auxi}). At the moment, let us limit ourselves to positive $z_1$ as stated in the main text (In Sec. IV below, we discuss the case with a negative $z_1$), Eqs. (\ref{eq:a1}) and (\ref{eq:a2}) lead to the following constraints on $y$:
\bea\label{eq:a3}
y^{-1} &\geqslant& \frac{z_1}{2}-1,~~\mathrm{for}~xz_1\geqslant 0,\nonumber\\
y^{-1} &\leqslant& -1,~~\mathrm{for}~xz_1< 0.
\eea
To have a non-negative efficiency at maximum power for $y$ satisfying Eq. (11) in the main text which guarantees a non-negative entropy production rate, the above inequalities imply that we should require 
\bea\label{eq:a4}
y^{-1} &\geqslant& h(x,z_1,z_2)~\geqslant~1-\frac{1}{z_2}~\geqslant~\frac{z_1}{2}-1,~~\mathrm{for}~xz_1\geqslant 0,\nonumber\\
y^{-1} &\leqslant& h(x,z_1,z_2)~\leqslant~-\frac{1}{z_2}~\leqslant~-1,~~\mathrm{for}~xz_1< 0.
\eea
It can be easily deduce that inequalities $1-\frac{1}{z_2}\geqslant\frac{z_1}{2}-1$ and $-\frac{1}{z_2}\leqslant-1$ in Eq. (\ref{eq:a4}) can be simultaneously satisfied when we limit to the following ranges of $z_{1,2}$ 
\begin{equation}\label{eq:a5}
0~\leqslant~z_1~\leqslant~2,~~\mathrm{and}~~\frac{2}{4-z_1}~\leqslant~z_2~\leqslant~1.
\end{equation}

\subsection{B. Conditions for non-negative maximum efficiency}
To have a non-negative maximum efficiency, it is enough to require $\Omega\geqslant 0$ [cf. Eq. (\ref{eq:Omega})] and $J_Q^+\geqslant 0$ [cf. Eq. (\ref{eq:JQ_pm})] by noting results in Sec. I. Firstly, we get 
\bea\label{eq:s19}
\Omega &=& 4\mathcal{F}_Q^2L_{QQ}\left[L_{QQ}(L_{WW}-L_{WA})^2-L_{WQ}(L_{WW}-L_{WA})(L_{QW}-L_{QA})\right]\nonumber\\
&=& \frac{4\mathcal{F}_Q^2L_{WQ}^4}{x^2}\frac{1}{z_1}\left(\frac{1}{y}+1\right)\left[\frac{1}{z_1}\left(\frac{1}{y}+1\right)-1\right]~\geqslant~0,\nonumber\\
\Rightarrow&&\frac{1}{z_1}\left(\frac{1}{y}+1\right)\left[\frac{1}{z_1}\left(\frac{1}{y}+1\right)-1\right]~\geqslant~0,\nonumber\\
\Rightarrow&& y^{-1}~\geqslant~z_1-1~~~\mathrm{or}~~~y^{-1}~\leqslant~-1
\eea
since we limit ourselves to positive $z_1$ (in Sec. IV, we discuss the case with a negative $z_1$), and
\bea\label{eq:s20}
&& J_Q^+\geqslant 0~\Rightarrow~L_{WW}-L_{WA}\geqslant 0\nonumber\\
&& \Rightarrow \frac{1}{xz_1}\left(\frac{1}{y}+1\right)\geqslant 0~\Rightarrow~\left\{\begin{array}{c}
y^{-1}\geqslant -1,~~\mathrm{for}~xz_1\geqslant 0,\\
y^{-1}\leqslant -1,~~\mathrm{for}~xz_1< 0.
\end{array}\right.
\eea
Therefore, to have $\Omega\geqslant 0$ and $J_Q^+\geqslant 0$ simultaneously, we should require
\begin{equation}\label{eq:c3}
\left\{\begin{array}{c}
y^{-1}\geqslant z_1-1,~~\mathrm{for}~xz_1\geqslant 0,\\
y^{-1}\leqslant -1,~~\mathrm{for}~xz_1< 0.
\end{array}\right.
\end{equation}

To enable a non-negative maximum efficiency for $y$ satisfying Eq. (11) in the main text, Eq. (\ref{eq:c3}) indicates that we should consider
\bea\label{eq:c4}
y^{-1} &\geqslant& h(x,z_1,z_2)~\geqslant~1-\frac{1}{z_2}~\geqslant~z_1-1,~~\mathrm{for}~xz_1\geqslant 0,\nonumber\\
y^{-1} &\leqslant& h(x,z_1,z_2)~\leqslant~-\frac{1}{z_2}~\leqslant~-1,~~\mathrm{for}~xz_1< 0.
\eea
Inequalities $1-\frac{1}{z_2}\geqslant z_1-1$ and $-\frac{1}{z_2}\leqslant-1$ in Eq. (\ref{eq:c4}) can be simultaneously satisfied if we limit to the following ranges of $z_{1,2}$
\begin{equation}\label{eq:c5}
0~\leqslant~z_1~\leqslant~1,~~\mathrm{and}~~\frac{1}{2-z_1}~\leqslant~z_2~\leqslant~1,
\end{equation}
Comparing Eq. (\ref{eq:a5}) with Eq. (\ref{eq:c5}) and noting that $2/(4-z_1)<1/(2-z_1)$ for positive $z_1$,  one should take Eq. (\ref{eq:c5}) as the overall constraints on $z_{1,2}$ [Eq. (14) in the main text] so as to ensure both a non-negative efficiency at maximum power and a non-negative maximum efficiency.

\section{III. Bounding optimal efficiencies}\label{sec:3}
We start from the expressions for the optimal figures of merit
\bea
\eta_{\mathrm{EMP}}(x,z_{1,2}) &\equiv& \frac{x^2z_1^2z_2\eta_c}{z_2(x+z_1)^2+2xz_1(2z_2-2-z_1z_2)},\label{eq:31}\\
\eta_{\mathrm{ME}}(x,z_{1,2}) &\equiv& \eta_c|x|\left(\sqrt{\left|\frac{1}{z_1}(h+1)\right|}-\sqrt{\left|\frac{1}{z_1}(h+1)-1\right|}\right)^2\nonumber\\
&=& \frac{\eta_c}{4z_1^2z_2}\left(\sqrt{\left|z_2(x+z_1)^2+4xz_1(z_2-1)\right|}-\sqrt{\left|z_2(x+z_1)^2+4xz_1(z_2-1-z_1z_2)\right|}\right)^2.\label{eq:32}
\eea
Here, $\eta_c$ denotes the Carnot efficiency, EMP (ME) is short for ``efficiency at maximum power" (``maximum efficiency"). We first look at $\eta_{\mathrm{EMP}}$. To enable a fair comparison between weak and strong-coupling cyclic QHEs, we introduce a rescaled parameter $x'=x/z_1$ and find from Eq. (\ref{eq:31}) that
\bea\label{eq:s26}
\eta_{\mathrm{EMP}}(x',z_{1,2}) &=& \eta_c\frac{x'^2z_1^2}{(x'+1)^2+2x'(2-2/z_2-z_1)}~\leqslant~\eta_c\frac{x'^2}{M_{\mathrm{min}}}\nonumber\\
\eea
Here, $M_{\mathrm{min}}$ denotes the minimum of the function $M=[(x'+1)^2+2x'(2-2/z_2-z_1)]/z_1^2$ by varying $z_{1,2}$.
Noting $-2\leqslant 2-2/z_2-z_1 \leqslant 0$ from Eq. (\ref{eq:c5}), we find $(x'+1)^2+2x'(2-2/z_2-z_1)\geqslant (x'+1)^2-4x'\geqslant 0$, hence $M_{\mathrm{min}}=\left.[(x'+1)^2+2x'(2-2/z_2-z_1)]\right|_{z_{1,2}=1}=x'^2+1$ when $1/z_1^2=1$ or, equivalently, $z_{1,2}=1$ from Eq. (\ref{eq:c5}). Therefore, we get
\begin{equation}\label{eq:s27}
\eta_{\mathrm{EMP}}(x',z_{1,2})~\leqslant~\eta_c\frac{x^2}{x'^2+1}~\equiv~\eta_{\mathrm{EMP}}^{\mathrm{weak}}(x').
\end{equation}

We then turn to $\eta_{\mathrm{ME}}$. Eq. (\ref{eq:32}) yields ($x'=x/z_1$)
\bea\label{eq:s28}
\eta_{\mathrm{ME}}(x',z_{1,2}) &=& \frac{\eta_c}{4}\left(\sqrt{\left|(x'+1)^2+4x'(1-1/z_2)\right|}-\sqrt{\left|(x'+1)^2+4x'(1-1/z_2-z_1)\right|}\right)^2\nonumber\\
&\leqslant& \frac{\eta_c}{4}\left(\sqrt{(x'+1)^2}-\sqrt{(x'-1)^2}\right)^2~\equiv~\eta_{\mathrm{ME}}^{\mathrm{weak}}(x').
\eea
In getting the second line, we have noted $1-1/z_2\leqslant 0$ and $1-1/z_2-z_1\geqslant -1$ since $1/(2-z_1)\leqslant z_2\leqslant 1$ [cf. Eq. (\ref{eq:c5})], and used the fact that $\sqrt{|(x'+1)^2+4x'K|}$ is an increasing function of $K$. Notably,
$\eta_{\mathrm{ME}}^{\mathrm{weak}}(x')=\eta_c x'^2$ ($\eta_c$) when $|x'|\leqslant 1$ ($|x'|\geqslant 1$).

\section{IV. The case with a negative $z_1$}
For results showed above and in the main text, we limited ourselves to a positive $z_1$ by noting the property $h(-x,-z_1,z_2)=h(x,z_1,z_2)$. In this section, we will discuss in detail the scenario permitting a negative $z_1$. Firstly, we note that the derivations showed in Sec. I and that for $\eta(P_{\mathrm{max}})$ presented in the main text are independent of the sign of $z_1$, hence the expressions for optimal efficiencies $\eta(P_{\mathrm{max}})$ and $\eta_{\mathrm{max}}$ and their bounds $\eta_{\mathrm{EMP}}$ and $\eta_{\mathrm{ME}}$ depicted in the main text remain the same under a negative $z_1$.

Next, let us determine the ranges of $z_{1,2}$ that guarantee the existence of non-negative optimal efficiencies. From the Sec. II A, one can easily check that inequalities $1-\frac{1}{z_2}\geqslant\frac{z_1}{2}-1$ and $-\frac{1}{z_2}\leqslant-1$ in Eq. (\ref{eq:a4}) can also be satisfied when
\begin{equation}\label{eq:s29}
z_1~<~0,~~\mathrm{and}~~\frac{2}{4-z_1}~\leqslant~z_2~\leqslant~1.
\end{equation}
Now we revisit the results depicted in the Sec. II B when a negative $z_1$ is allowed. Eq. (\ref{eq:s19}) now yields
\begin{equation}
y^{-1}~\leqslant~z_1-1~~~\mathrm{or}~~~y^{-1}~\geqslant~-1.
\end{equation}
The above conditions together with Eq. (\ref{eq:s20}) which is unaltered indicate that Eq. (\ref{eq:c3}) should be replaced by
\begin{equation}\label{eq:s31}
\left\{\begin{array}{c}
y^{-1}\geqslant -1,~~\mathrm{for}~xz_1\geqslant 0,\\
y^{-1}\leqslant z_1-1,~~\mathrm{for}~xz_1< 0
\end{array}\right.
\end{equation}
for a negative $z_1$. Hence instead of Eq. (\ref{eq:c4}), we now should require
\bea\label{eq:s32}
y^{-1} &\geqslant& h(x,z_1,z_2)~\geqslant~1-\frac{1}{z_2}~\geqslant~-1,~~\mathrm{for}~xz_1\geqslant 0,\nonumber\\
y^{-1} &\leqslant& h(x,z_1,z_2)~\leqslant~-\frac{1}{z_2}~\leqslant~z_1-1,~~\mathrm{for}~xz_1< 0.
\eea
Inequalities $1-\frac{1}{z_2}\geqslant -1$ and $-\frac{1}{z_2}\leqslant z_1-1$ in Eq. (\ref{eq:s32}) can be simultaneously satisfied if we limit to the following ranges of $z_{1,2}$
\begin{equation}\label{eq:s33}
-1~\leqslant~z_1~<~0,~~\mathrm{and}~~\frac{1}{2}~\leqslant~z_2~\leqslant~\frac{1}{1-z_1}.
\end{equation}
Noting $2/(4-z_1)<1/2$ and $1/(1-z_1)<1$ by comparing Eqs. (\ref{eq:s29}) and (\ref{eq:s33}), we can infer that Eq. (\ref{eq:s33}) gives the required ranges for $z_{1,2}$ that permit non-negative optimal efficiencies for the scenario with a negative $z_1$. 

As the last step, we revisit the results presented in Sec. III by using Eq. (\ref{eq:s33}). We still have $-2\leqslant 2-2/z_2-z_1<0$ according to Eq. (\ref{eq:s33}) such that the function $M$ in Eq. (\ref{eq:s26}) attains its minimum when $1/z_2^2=1$, or equivalently, $z_1=-1$ and $z_2=1/2$ as can be inferred from Eq. (\ref{eq:s33}). Hence we still find $M_{\mathrm{min}}=\left.[(x'+1)^2+2x'(2-2/z_2-z_1)]\right|_{z_{1}=-1,z_{2}=1/2}=x'^2+1$ and Eq. (\ref{eq:s27}) is respected. Similarly, we still have Eq. (\ref{eq:s28}) since $1-1/z_2\leqslant0$ and $1-1/z_2-z_1\geqslant -1$ as well from Eq. (\ref{eq:s33}). As a result, we still have ($n=\mathrm{EMP}, \mathrm{ME}$)
\begin{equation}\label{eq:s34}
\eta_n^{\infty}~=~z_1^2\eta_c~~\mathrm{and}~~\eta_{n}~\leqslant~\eta_{n}^{\mathrm{weak}}
\end{equation}
even if a negative $z_1$ is allowed. In Fig. (\ref{fig:s1}), we present a set of results with varying $z_{1,2}$ taken according to Eq. (\ref{eq:s33}). As can be seen from the results shown in the figure, Eq. (\ref{eq:s34}) is clearly validated.
%
%-------------------------------------------------
\begin{figure}[thb!]
 \centering
\includegraphics[width=0.8\columnwidth]{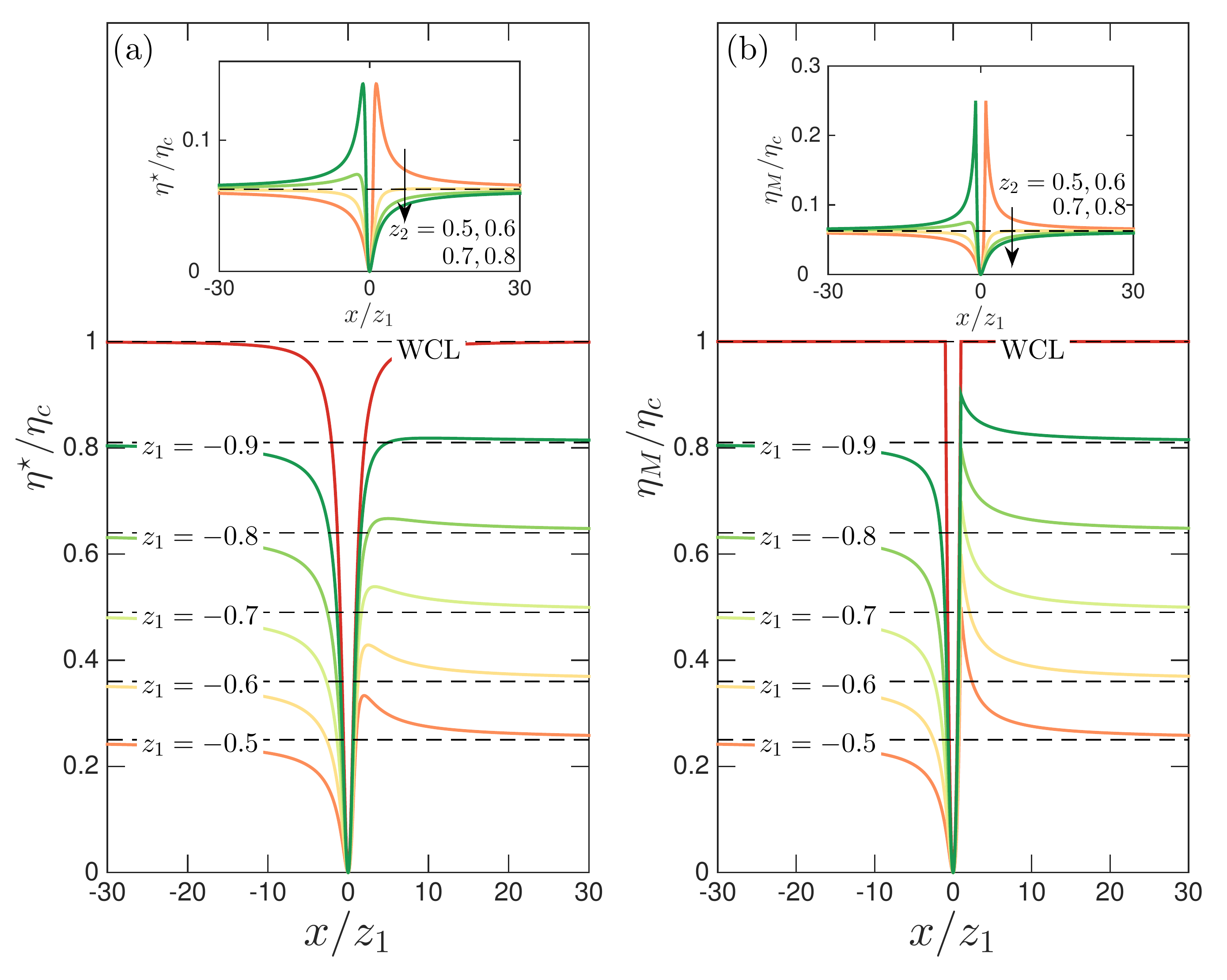} 
%\vspace*{-16mm}
\caption{(a) $\eta_{\mathrm{EMP}}/\eta_c$ [cf. Eq. (\ref{eq:31})] as a function of $x/z_1$ with varying $z_1$ and fixed $z_2=0.5$. Inset: Results with varying $z_2$ and fixed $z_1=-0.25$. (b) $\eta_{\mathrm{ME}}/\eta_c$ [cf. Eq. (\ref{eq:32})] as a function of $x/z_1$ with varying $z_1$ and fixed $z_2=0.5$. Inset: Results with varying $z_2$ and fixed $z_1=-0.25$. For comparisons, we depict the weak coupling limit (WCL) with $z_{1,2}=1$. Dashed horizontal lines in both plots mark the value of $z_1^2$.
}
\protect\label{fig:s1}
\end{figure}
%-------------------------------------------------

\end{widetext}

\end{document}